\documentclass[]{sig-alternate}

\conferenceinfo{ICSE}{'14, May 31 - June 7, 2014, Hyderabad, India}
\CopyrightYear{14}
\crdata{978-1-4503-2756-5/14/05}

\usepackage[colorlinks,linkcolor=black,citecolor=black]{hyperref}
\usepackage[utf8]{inputenc}
\usepackage{listings}
\usepackage{multicol}
\usepackage{framed}
\usepackage{tabularx}
\usepackage{balance}
\usepackage{textpos}
\usepackage{xcolor}
\usepackage{graphicx}

\begin{document}
\title{A Critical Review of ``Automatic Patch Generation Learned from Human-Written Patches'': Essay on the Problem Statement and the Evaluation of Automatic Software Repair}
\author{Martin Monperrus\\
\affaddr University of Lille \& INRIA, France\\
\affaddr martin.monperrus@univ-lille1.fr\\
}

\maketitle

\hypersetup{
pdftitle={A Critical Review of "Automatic Patch Generation Learned from Human-Written Patches'": Essay on the Problem Statement and the Evaluation of Automatic Software Repair},
pdfauthor={Martin Monperrus},
pdfproducer={},
pdfcreator={}
}

\begin{textblock*}{20cm}(-1cm,-1cm)
\begin{center}
\end{center}
\end{textblock*}

\newcommand\wsection[1]{}
\newcommand\xsection[1]{\section{#1}}
\newcommand\ysection[1]{\subsection{#1}}

\newcommand\xparagraph[1]{\paragraph{#1}}

\newenvironment{newtext}{}{}

\wsection{Part 1: A Critical Reading of ``Automatic Patch Generation Learned from Human-Written Patches``}

\begin{abstract}
At ICSE'2013, there was the first session ever dedicated to automatic program repair. In this session, Kim et al. presented PAR, a novel template-based approach for fixing Java bugs. We strongly disagree with key points of this paper. Our critical review has two goals. First, we aim at explaining why we disagree with Kim and colleagues and why the reasons behind this disagreement are important for research on automatic software repair in general. Second, we aim at contributing to the field with a clarification of the essential ideas behind automatic software repair. In particular we discuss the main evaluation criteria of automatic software repair: understandability, correctness and completeness. We show that depending on how one sets up the repair scenario, the evaluation goals may be contradictory. Eventually, we discuss the nature of fix acceptability and its relation to the notion of software correctness.
\end{abstract}



\section{Introduction}

\begin{newtext}

\begin{quotation}
``What one would like ideally [...] is the
 automatic detection and correction of bugs'' R. J. Abbott, 1990 \cite{abbott1990resourceful}
\end{quotation}

The automatic detection of bugs has been a vast research field for decades, with a large spectrum of static and dynamic techniques. 
Active research on the automatic repair\footnote{or ``correction'', ``fixing'', ``patching'', \ldots} of bugs is more recent. 
A seminal line of research started in 2009 with the GenProg system \cite{Weimer2009,Goues2012}, and at the 2013 International Conference on Software Engineering, there was the first session ever dedicated to automatic program repair.

The PAR system \cite{Kim2013} was presented there, it is an approach for automatically fixing bugs of Java code.
The repair problem statement is the same as GenProg \cite{Goues2012} ``given a test suite with at least one failing test, generate a patch that makes all test cases passing''. 
PAR introduces a new technique to fix bugs, based on templates.
Each of PAR's ten repair templates represents a common way to fix a common kind of bug. 
For instance, a common bug is the access to a null pointer, and a common fix of this bug is to add a nullness check just before the undesired access: this is template ``Null Pointer Checker''.

We strongly disagree with Kim et al.'s paper on PAR.
This is our motivation to present this critical review of their work. 
We think that a respectful critical reading and debate is very important in the scientific process. 
This usually happens within technical contributions (e.g. in the related work section), but this is often shallow and biased towards the new approach.
Explicit criticism may sometimes be better to clearly see the opposing ideas in the first place (see for instance the strong opposition in ``A Debate on Teaching Computer Science'' in the Communications of the ACM \cite{Denning1989}).

Our critical review has two goals.
First, we aim at explaining why we disagree with Kim and colleagues and why the reasons behind this disagreement are important for automatic software repair in general.
Second, we aim at contributing to the field with a clarification of the essential ideas behind automatic software repair.
We will discuss neither the motivation of automatic software repair (we extensively work on automatic repair) nor the technical contribution (we believe  in the efficiency of templates for fixing certain bugs and the soundness of PAR's ones).

We start by discussing the concept of ``defect class'' which is missing in PAR's paper. 
We show that ignoring this concept has serious consequences on the conclusiveness of evaluation protocols in general, and PAR's one in particular.
We propose three dimensions for defining a defect class: the root cause, the symptom, and the kind of fix. This is broad enough to revisit in the related work recent papers on automatic repair.

Then, we will elaborate on the problem statement of automatic repair.
Beyond the canonical problem statement \cite{Weimer2009,Goues2012,Kim2013} which we call test-suite based program repair, we show that there are diverse repair problems, depending on whether repair happens online at runtime or offline at maintenance time, and whether repair happens on the state of programs or on the behavior expressed in their code. 

Naturally, this diversity of problem statements leads to a diversity of evaluation criteria. 
For test-suite based program repair, we emphasize on the need to characterize the inner quality of test suites.
Then, depending on whether one considers software repair as fully automatic code synthesis or as a recommendation system problem, there are different evaluation criteria to be considered. Those evaluation criteria may even be contradictory.
For instance, bug fix recommendation systems are expected to synthesize understandable patches but are allowed to provide partial solutions. Fully automatic repair systems have no understandability constraints but should always yield a 100\% executable solution.

Eventually, we explore the foundations of the fix acceptability question (whether a fix is more acceptable than another).
Through a thought experiment, we claim that under certain circumstances, the question ``is patch A more acceptable than patch B?'' is fundamentally unanswerable. We finish the discussion by revisiting the relation between repair and software correctness.

To sum up,  we contribute to the field of automatic repair with new perspectives using Kim et al.'s paper as starting point.
Our contributions are:
\begin{itemize}
\item the highlighting of pitfalls in automatic software repair research and the uttermost importance of explicit defect classes;
\item the presentation of kinds of software repair and their characteristics, in particular state repair and behavioral repair;
\item the identification of important and sometimes contradictory evaluation criteria in the field: understandability; correctness; completeness.
\end{itemize}

The paper reads as follows.
Section \ref{sec:background} gives some background information on automatic software repair and PAR.
Section \ref{sec:wrong} is the critical review per se, it exposes what we think is wrong in Kim et al.'s paper.
Section \ref{sec:problem-statement} elaborates on the problem statement of automatic software repair.
Section \ref{sec:eval} discusses the evaluation of automatic software repair techniques and presents a thought experiment on the notion of ``fix acceptability''.
Section \ref{sec:rw} discusses the related work. 
Section \ref{sec:conclusion} concludes this paper.

\end{newtext}

\begin{newtext}

\section{Background}
\label{sec:background}

Before discussing in details Kim et al.'s approach called PAR \cite{Kim2013}, let us first briefly discuss what we mean by automatic software repair and then how PAR works. 

\subsection{Automatic Software Repair}
\label{sec:background-asr}
Automatic software repair is the process of fixing software bugs automatically. 
This definition seems broad enough to encompass diverse approaches from different fields of computer science and engineering.
In particular, it accepts different notions of fixing (e.g. on source code, on binary code, on execution data) and different kinds of bugs (independently of the severity and the type of oracle that asserts the presence of the bug). In a nutshell, automatic software repair consists of several overlapping phases: failure detection (something wrong has happened), bug diagnosis (why this has happened), fault localization (what the root cause is, where the initial faulty module/statement is), repair inference (what should be done to fix the problem). 
For instance, an archetypal automatic software repair system takes a bug report as input and outputs a source code patch that fixes the bug.

It closely relates to what is called ``automatic debugging'' \cite{zeller2001automated}: the ultimate goal of debugging is to fix bugs and to that extent automatic software repair is perfect automatic debugging. However, as said above, the term ``automatic software repair'' is broader in scope than automatic debugging in the sense that debugging connotes more the diagnosis phase of repair than the combination of the aforementioned phases. Automatic software repair relates to the traditional field of software dependability \cite{avizienis2004basic}, and in its terminology, it spans both fault tolerance and fault removal. Indeed, automatic software repair can benefit from the clear definitions of ``error'', ``failure'', and ``fault'' coming from dependable computing  \cite{avizienis2004basic}, because the term ``bug'' is vague and refers to those three concepts indistinctly.  

Automatic software repair also much relates to software testing (w.r.t. the techniques used to detect and diagnose the bugs), 
to  program synthesis (when it comes to create a correct piece of code), 
and to data mining and machine learning for software engineering (when the repair knowledge is extracted from existing data -- version control systems, execution traces, etc.). This list of related fields is not meant to be exhaustive.

\subsection{PAR: Template-based Software Repair}
\label{sec:background-par}

PAR is an automatic software repair technique invented by Kim and colleagues and presented at the 2013 International Conference on Software Engineering \cite{Kim2013}. 
As previous work \cite{Goues2012}, it states the problem of automatic repair as follows:
a bug is detected by a failing test case, and the goal is to change the application code so that, first the failing test case now passes and second, the other test cases (forming a ``test suite'') still pass.

To do so, PAR uses an existing fault localization technique, and for each suspicious statement, it tries different repair templates.
If the application of a repair template makes the test suite passing, the bug is considered to be fixed. 
Each of the ten repair templates represents a common way to fix a common kind of bug. 
For instance, a common bug is the access to a null pointer, and a common fix of this bug is to add a nullness check just before the undesired access (template ``Null Pointer Checker''). Some of PAR's templates use the intrinsic redundancy of software and pick some code elsewhere in the program under repair to fix the bug \cite{Martinez2014}. 

\end{newtext}

\section{A Critical Review of PAR} 
\label{sec:wrong}
\begin{table*}
\caption{Defect classes can be defined along several dimensions. Automatic repair approaches can be compared only if they address similar defect classes.}
\vspace{.2cm} 	
\begin{tabularx}{\textwidth}{p{4.5cm}|X}
\textbf{Defect class according to\ldots} & \textbf{Examples of defect class}\\
\hline
The root cause & Incorrect variable initialization, incorrect configuration, \ldots \\ \hline
The symptom & Segmentation faults, Null pointer exceptions, memory exhaustion, \ldots \\ \hline
The fix & Adding an input check, changing a method call, restoring an invariant, \ldots \\  
\end{tabularx}
\label{tab:defect-classes}
\end{table*}

In our group, we extensively work on automatic software repair \cite{Martinez2013,Martinez2014,demarco2014}.
In the last months, we have spent much time in studying Kim et al.'s paper on PAR \cite{Kim2013} and its dataset.
Over weeks, we have isolated two points on which we strongly disagree.
Beyond PAR only, this discussion aims at being insightful for automatic software repair as a whole.

\subsection{What are the addressed defect classes?}
By ``defect class'', we mean a family of bugs that have something in common (we equate ``bug'' and ``defect'', the former being the colloquial name for the latter).
We have an open understanding of ``in common'': they can share the same root cause (e.g. a programmer mistake), the same symptom (e.g.. an exception) or the same kind of fix (e.g. changing the conditional of an if expression).
Table \ref{tab:defect-classes} gives examples of such defined defect classes. There is often a relation between the root cause and the kind of fix. For instance, an incorrect initialization can obviously be fixed changing the initialization. But it can be also be fixed by inserting an assignment later. There can be several different fix locations (e.g. at line 42 or at line 666) and kinds for the same root cause (e.g. changing an assignment or changing a return value).
The core of PAR is a collection of 10 templates.
What are the addressed defect classes?

PAR doesn't address or clearly  identify one or several defect classes.
By looking at the template names and descriptions, they seem to address many different classes (a null pointer bug is of different nature compared to a incorrect cast bug). 
There is no apparent principle behind the collection of templates. 
As far as we understand, the templates seem to have been collected by browsing bug fixes and see whether they would fit in their overall approach. Why is this ad hoc approach a problem? 

To us, a contribution on automatic software repair should answer the following questions:
For which defect class does it work? 
This names the enemy and enables the community to answer the related questions: what are the ``repairable'' defect classes, why is a defect class easy/hard to repair? 
Identifying a defect class in automatic repair is as important as defining the ``fault model'' \cite{avizienis2004basic} of a testing technique\footnote{Our goal is not to articulate those two terms. For us, a fault model consists of several defect classes and may identifiy some relations between the root causes and the symptoms.}.

\begin{framed} 
Identifying the target defect classes enables the community to answer the questions: what are the ``repairable'' defect classes, why is a defect class easy or hard to repair? 
\end{framed} 

These are academic questions. The real impact of automatic repair on practitioners is founded are two other questions:
What's the abundance of this defect class? What is its criticality?
Automatically fixing a common defect class would save a lot of maintenance resources, fixing critical crashes  would save a lot of production loss.
On the contrary, the automatic repair of minor and rare bugs would only be of academic interest.
The lack of explicit defect classes in PAR's paper hinders the answering to all those questions.

Note that the seminal paper on automatic repair by Weimer and colleague (GenProg, \cite{Weimer2009}) does not explicitly address those questions as well. However, it was a paper opening a new field of research. But we note that in the conference presentation \footnote{Slide 20 of \url{http://dijkstra.cs.virginia.edu/genprog/papers/weimer-icse2009-genprog-presentation.pdf}} and in subsequent papers \cite{Goues2012}, the notion of defect class is clearly present and hints that GenProg works best for manipulating defensive code against memory errors (in particular segmentation faults and buffer overruns).

\subsection{How conclusive is the evaluation?}

Let us now discuss the conclusiveness of PAR's evaluation \cite{Kim2013}. 
Using 119 real bugs, they evaluate whether PAR fixes more bugs than GenProg \cite{Goues2012} and whether the generated fixes are of better quality.
We think that the experimental methodology has several issues.
Note that we are not saying that their approach does not work (actually, we think that their AST template-based approach may well address some common bugs).
We are only saying that the conclusiveness of the evaluation as presented is questionable.

\subsubsection{On the relation between the dataset creation methodology and conclusiveness} 
\label{sec:dataset-fallacy}
The first part of the evaluation is about the number of fixed bugs.
In PAR's paper, there are 10 templates, and in the presented evaluation, PAR is able to fix 27 bugs.
What is the distribution of fixed bugs by template?
Mathematically, the templates and their associated defect class are evaluated on average on 2 or 3 bug instances ($27/10=2.7$).
We asked the authors about this distribution (i.e. which template fixes which bug ID?): it has been lost.
According to our replication experiments with the 27 fixed bugs, it actually seems that most bugs are fixed by the same templates (in particular ``Null Pointer Checker'' and ``Expression Adder, Remover, Replacer''), and that others, e.g. ``Class cast checker'', only fix one bug.
 
Many empirical evaluations, including some of ours, are biased or over-conclusive with the proposed approach.
The bias often lies in the way the datasets are constructed.
However, beyond the magnitude of a validation, the dataset construction can impact the internal validity.

Do PAR and GenProg address the same defect classes?
If no, this raises doubts about the conclusiveness of the results. Let's assume that PAR address defect classes A and B and GenProg defect classes B and C. 
In this case, depending on how one builds the dataset, the results would be totally different: for instance, a dataset 80\%(A)-20\%(B) would much favor PAR.
This fallacy is partly due to the absence of the concept of defect class.
In PAR's paper, there is no presentation on how the dataset was built and no characterization of the kind of bugs it contains.
To address this fallacy, one needs to characterize how the dataset is built and what it contains.

\begin{framed}
The way one builds evaluation datasets for automatic repair has a great impact on the conclusiveness of the results.
\end{framed}

Finally, is PAR better than GenProg? According to our arguments, as long as one does not clarify the underlying defect classes and build a well-formed dataset, we do not know. If PAR and GenProg  are proven to address different defect classes, the question can be considered as ill-formed.

\begin{newtext}
The question that then arises is: how to build a valid evaluation dataset for automatic repair?
At this point in the life of the research field, there is no definitive answer.
However, we tend to think that having a explicit target defect class is again a key to answering this question. 
The dataset should contain only bugs from the same defect class.
Within a defect class, the dataset sampling should be stratified from easy bugs to complex ones. 
With such a dataset, subsequent approaches on the same defect class can be meaningfully compared.
Having yet-unfixed complex bugs in the dataset would even foster creativity in other teams, who would invent new ways to fix them in an automated manner.
\end{newtext}

\begin{table*}
\caption{Two examples of software repair scenarios. Depending on the problem, the evaluation criteria of automatic software repair are different and the evaluation goals may be contradictory.}
\vspace{.2cm}
\begin{tabularx}{\textwidth}{p{2.8cm}|p{6cm}|X}
\textbf{Facet} & \textbf{Fully Automatic System} & \textbf{Recommendation System} \\
\hline
Who & The repair robot or agent & The repair system, then the human\\
When & Mostly at runtime & Mostly at development and maintenance time\\ 
Longevity & Solutions may be temporary, disposable & Solutions should sustain time \\
\hline
Understandability & Anything goes incl. alien code  & Solutions must be understandable by humans \\
Correctness & Fully automated procedure & Pre-filtering then human validation \\
Completeness & Solutions must be 100\% executable & Solutions may be partial, a human would fill the gaps\\
\end{tabularx}
\label{tab:repair-vs-recommendation}
\end{table*}

\subsubsection{On the meaning of evaluating ``patch acceptability''} 
\label{sec:acceptability}
Let us now examine the second evaluation question on whether the synthesized patches are of better quality compared to ones synthesized with an other approach, or compared to the real ones. The experimental protocol consists of asking developers to blindly assess synthesized patches. They are also asked to assess the original human-written patch. The experiment says it evaluates ``patch acceptability''.
The subjects of the user study are 17 students and 67 developers of local companies. None of them are developers of the software packages for which bugs are fixed.
As a result of this evaluation, it is claimed that ``PAR generates more acceptable patches than GenProg does''. What are the underlying assumptions of this experiment?

First, it is that a developer is able to rate the quality of a patch without any knowledge about the codebase and domain of the bug. 
Within reasonable time in the experiment (say 30 minutes per bug), we think it is hardly possible.
For many bugs, understanding the bug report itself, understanding the causality chain\footnote{This often requires executing the program and not only looking at the program as done in this setup.}, and understanding whether the patch is correct are all difficult tasks. Given only the patch and an hyperlink to the bug report, doing this analysis on an unknown codebase seems really hard. Understanding the inner quality of a patch requires far more domain-specific knowledge than the subjects of the user study have.
To us, the subjects of the experiment do not rate the inner quality of the patch, but more on whether the code ``looks good' or not (Kim et al. use ``is acceptable'', but as explained above, to us, we think that ``looks good'' fits more the reality of the experiment).

\begin{framed}
Evaluating the inner quality of a patch requires a thorough process involving understanding the bug report itself, understanding the causality chain of the bug, and understanding the potential consequences and side-effects of the patch.
\end{framed}

\section{On the Problem Statement of Automatic Software Repair}
\label{sec:problem-statement}
Now, let us step back and discuss the core problem statement of the field of automatic software repair. 

\subsection{On Patch Prettiness versus Alien Code}

As said above and especially in Section~\ref{sec:acceptability}, to our understanding, PAR's evaluation \emph{implicitly} reformulates the automatic repair problem statement as ``generate a patch that makes the test suite passing and that looks good''.
Beyond ``looks good'', we would say ``looks like humanly written good code''. 

Automatic software repair is one branch of code synthesis: according to the literature and our experience, code synthesis often generates surprising code, a kind of alien code. This is normal since the processes to create this are completely different  (biological versus artificial).
Let us dwell on this. Automatic software repair is about fixing bugs automatically, it is not about fixing bugs ``as humans are used to''. We should not be afraid of alien ways of reasoning on and modifying programs (alien in the sense of fundamentally different). 

PAR's implicit reformulation of the problem statement of automatic repair puts up barriers on the way we fix bugs and on the way we design automatic bug fixing techniques.
We should not only aim at techniques that mimic human bug fixing. 

\begin{framed}
We should not be afraid of alien ways of writing code.
\end{framed}

\subsection{Kinds of Software Repair}
\begin{newtext}
More generally, the problem statements of automatic software repair are not yet clearly identified. 
In the following, we first aim at clarifying the ``canonical'' repair problem, as stated by Weimer et al.  \cite{Weimer2009}, and we call it ``test-suite based program repair'' (Section \ref{sec:ts-repair}).
We then broaden the scope of automatic repair and propose to distinguish two families of repair techniques:
state repair and behavioral repair  (Section \ref{sec:state-vs-behavioral}).
\end{newtext}

\subsubsection{Test-suite based Program Repair}
\label{sec:ts-repair}

In both PAR \cite{Kim2013} and GenProg \cite{Weimer2009}, the primary problem statement is ``given a test suite with one failing test, generate a patch that makes them all passing''.  The failing test case is the oracle for the bug.
The rest of the test suite is the oracle for regression.
This is what can be called ``test-suite based program repair''.
In the patch quality experiment of PAR, all the assessed patches pass the whole test suite. It means that the primary problem statement is solved.

Asking users whether a patch is better than another one implicitly breaks the well-formedness of the problem statement. Asking the question means that passing the test suite may not be sufficient. 
Many have raised such points since GenProg's break-up.
With hindsight, we also agree that this is an important weakness of the original problem statement as done by Weimer and colleagues.
However, as shown above, the answer of PAR yields an implicit reformulation of the automatic repair problem statement as ``generate a patch that makes the test suite passing and that looks good''. We disagree with it. We think that a better way to reformulate it is to put the emphasis on assessing the test suite quality: to what extent is a test suite good? does it well specify its domain? is it appropriate for automatic repair?

\begin{framed}
If the research community is able to characterize what a good test suite is, we can simply clarify the problem statement as follows, ``given a good and trustable test suite, generate a patch that makes the test suite passing''.
\end{framed}

It may happen that the concept of test suite as we understand it today -- with test cases and assertions -- will never prove appropriate for automatic repair.

More generally, in test suite based program repair, there is an important asymmetry between test cases. One single failing test case is enough to express the bug, while the other ones must cover as completely as possible the specified behavioral space. To this extent, the point ``given a good and trustable test suite'' is simply a concrete instance of the more general  problem of being able to characterize the specification of a program or software component \cite{staats2011programs}.

\begin{newtext}

\subsubsection{Behavioral Repair versus State Repair}
\label{sec:state-vs-behavioral}

GenProg \cite{Weimer2009} focuses on synthesizing code to fix bugs offline, the code being meant to be committed into a version control system. The visibility and impact of this work tends to associate the term ``repair'' with this kind of repair.
However, ``repair'' has a broader sense than just synthesizing source code.
It may mean repairing a data structure \cite{demsky2003automatic}, repairing the register values and memory locations \cite{Perkins2009}, etc.

We think that we can actually distinguish two kinds of automatic software repair: state repair and behavioral repair.
\emph{State repair} consists in modifying the program state during the execution (the registers, the heap, the stack, etc.).
Demsky and Rinard's work on data structure repair \cite{demsky2003automatic} is an example of such state repair.
State repair can be seen as a kind of data repair (as opposed to code repair), in the spirit of the data diversity techniques in fault tolerance \cite{ammann88}.
\emph{Behavioral repair} consists in modifying the program executable code. According to this definition, synthesizing a source code patch is indeed behavioral repair. Behavioral repair is also relevant on binary code when no source code is available. Behavioral repair can also happen at runtime (when one changes the code part of the memory).
To sum up, state repair is only online (at runtime) while behavioral repair can be either online or offline at maintenance time. We will come back on this point in Section \ref{sec:eval}.

\begin{framed}
The problem statement of automatic software repair can be decomposed in:
state repair that consists in modifying the program state during the execution, 
and behavioral repair that consists in modifying the program code.
\end{framed}

\end{newtext}

\section{On the Evaluation of Automatic Software Repair}
\label{sec:eval}

\begin{newtext}

According to the broad definition given in Section~\ref{sec:background-asr}, 
automatic software repair can be declined in different scenarios.
Repair involves failure detection, bug diagnosis, fault localization and repair inference.
Even though automatic software repair contains the word ``automatic'', 
it is not reasonable to state that automatic software repair systems are only those that fully automatically cover all those phases. 
A system that produces a patch on which a developer would build on to write the final patch indeed goes in the direction of automatic software repair. 
One can actually imagine a broad range of repair scenarios:
fully automatic repair agents at runtime,
repair bots taking care of some bug reports and modifying the source code base automatically,
or repair recommendation systems proposing tentative patches that developers would improve. 
As we shall see now, depending on the scenario, the evaluation criteria change and may even be contradictory.

\subsection{Evaluation Criteria}
\label{sec:automatic-recommendation}
We see at least three dimensions of evaluation for which the evaluation goals differ depending on the automatic repair scenario.

\xparagraph{Understandability}
Let us consider a repair robot  that automatically commits patches to a code base where conventional human-based maintenance takes place.
Some complex bugs would be unfixable by the repair robot and they would consequently be handled by a human developer. 
To fix those tough bugs, she may have to understand a patch previously generated by the repair robot, or even a set of superimposed synthesized patches, which may be rather difficult.
When human-based maintenance and automatic repair are interleaved, the generated patches have to be clearly documented, and the repair approach could also generate an explanation of the repair. Le Goues et al. refer to this issue as ``patch maintainability'' \cite{Goues2013}.
On the contrary, if the repair happens at runtime as a temporary solution in order not to crash, there is no need for documentation at all. 
In the former case, alien code is problematic, in the latter code, alien code is welcome.
State repair at runtime have no understandability requirements, and depending of the scenario of behavioral repair, understandability is not mandatory.
\emph{Synthesizing maintainable repairs is antagonist to runtime disposable fixes.}

\xparagraph{Correctness}
Those different repair scenarios have a direct impact on the correctness evaluation criteria.
To some extent, a repair system that generates patches is a recommendation system for software engineering.
In the mindset of using a recommendation system, the developer would use the system as follows.
She would consider a synthesized patch, perform additional correctness and understandability assessment, and then decide for the final patch to be committed to the repository. 
In such a scenario, the repair system is allowed to synthesize partially correct patches.
On the contrary, a fully automatic repair system is liable for synthesizing fully correct repairs\footnote{
Interestingly, there is asymmetry between state repair and behavioral repair with respect to correctness:
assessing the correctness of a new state (inferred, synthesized) corresponds to assessing one point;
on the contrary, assessing the correctness of a behavioral patch  requires to assess the correctness of the path under all anticipated inputs (many points).} (according to its correctness oracle).
The latter has a much larger scope and seems more difficult in general. 
\emph{The value of the repair system may come either from the fully automated correctness decision procedure or from the help it provides to the developers.}

\xparagraph{Completeness}
The same argument applies to the completeness of the patch.
A repair recommendation system can only provide a partial repair (say 90\% of the final repair), or even a sketch of the repair.
This can nonetheless be very valuable to guide the developer in writing the final patch. 
On the other hand, at runtime, the repair must be executable, and partialness is not an option. 
\emph{Partial repair is sometimes a valuable option and in other cases an unacceptable solution.}

\begin{framed}
Depending on the repair problem statement, the evaluation criteria are different and the evaluation goals may even be contradictory.
\end{framed}

Table \ref{tab:repair-vs-recommendation} sums up those points.  
For instance, bug fix recommendation systems are expected to synthesize understandable patches but are allowed to provide partial solutions. Fully automatic repair systems have no understandability constraints but should always yield a 100\% executable solution.

\end{newtext}

\subsection{On Fix Acceptability}
\label{sec:fix-acceptability}
Let us consider again the question of fix acceptability. 
We now go beyond  ``looks good and humanly written'', and more generally beyond understandability, correctness and completeness.
Beyond those three evaluation criteria, we think that the question of fix acceptability is related to the foundations of software.
Let us assume that we have a good and trustable test suite.
This test suite completely specifies the expected behavior in the sense of ``if the repair technique mixes up the software'' then ``the introduced bugs will be detected''.

\begin{lstlisting}[float,caption=Two possible fixes of the same bug (both satisfy all test cases). There is no unique way to say that one is more acceptable than the other.,label=lst:the-problem,captionpos=b]
// fix A: code insertion at line 21
+ if (x==2) { foo(x); }

// fix B: code insertion at line 21
+ if (x<=2) { foo(x); }
\end{lstlisting}

Now, let us assume that for a given bug and its failing test case, one has two possible fixes which are shown in Listing \ref{lst:the-problem}.
Obviously, both patches fix a test case where the value ``2'' is involved. This value has a semantics in its domain.
According to the setup, either foo is idempotent for all values $<2$ or no values $<2$ are tested.

In the latter case, since we assume one has a good test suite, it means that for values $<2$ , the behavior is unspecified.
This can also be formulated as the values $<2$ are outside the specified domain (neither a nominal value nor an expected incorrect value). In both cases, there is no direct answer to the question ``which patch is more acceptable than the other''.

Let us now consider the topology of the output domain. The first fix A (with ``\texttt{==}'') does nothing outside the specified point of the failing test case. The second fix B (with ``with \texttt{<=}'') has an impact on the behavior within the unspecified input domain.
Some tend to prefer fix A because it minimizes the impact which is a well-known engineering value. 

However, the first fix A (with ``\texttt{==}'') introduces an irregularity: something happens only for one point of the input domain.
On the contrary the second fix B (with ``with \texttt{<=}'')  is more ``regular'', it introduces a kind of phase transition at $x=2$.
This fits more to the idea that $x=2$ does not represent an exception, but a boundary, which often happens in input domains.
Some tend to prefer fix B because it minimizes the number of irregularities which is also a well-known engineering value.

Even within our research group, there is no consensus on which patch is better.
What we want to show with this made-up example is two fold. 
First, the notion of ``fix acceptability'' is actually founded on deep concepts and beliefs on the nature of software. It may depend on the domain itself.
Second, there may be many concurrent fixes for which there is simply no answer. In other terms, asking the question ``which patch is more acceptable than the other'' must be done with great care, by explicitly stating that ``None'' is a valid and common answer. 

This has been done in PAR's experiment: the subjects could answer ``both [patches] are acceptable''. However, as discussed above,  the subjects' absence of knowledge about the domain tends to show that when the subjects answered this, this had little to do with answering ``the patches are incommensurable''.

\begin{framed}
Fix acceptability may be an unanswerable question.
\end{framed}

\begin{newtext}
It is now clear that the question of fix acceptability is directly related to what can be considered as correct or not, directly related to the nature of software correctness.
Program repair lies at the conjunction of two dimensions of software correctness.
It needs an oracle of what is incorrect: an oracle for the bug.
It also needs an oracle on what behavior should be kept correct for sake of non-regression.

The conventional, common sense, notion of software correctness is binary:
there is a decision procedure that says whether the software is correct (the procedure outputs ``true'') or not (the procedure outputs ``false'').
Dijkstra says \emph{``a program with an error is just wrong''} \cite{Denning1989}.
Boolean assertions in programming languages and testing frameworks embody this notion.
In this perspective, the problem statement of program repair is easy: the binary oracle of the bug should be negated and the other binary oracles should be kept passing. This is the ``canonical'' problem statement of program repair, as stated by GenProg.

However, binary software correctness is no longer the norm.
Software correctness exists at different scales: e.g. at the level of expressions (e.g. arithmetic expressions), functions, modules, systems, etc.
Let us differentiate between the two extremes as ``local correctness'' (a few lines of code) and ``global correctness'' (at the system level, up to several millions of lines of code).  
Small-scale correctness is often binary.  However, when the scale increases, a different kind of correctness emerges.
This emerging correctness offers two new facets:
it may be partial as when  a system passes 1990 test cases out of 2000;
and it may be continuous, as when one considers the quality of service \cite{misailovic2010quality} (many quality of service attributes are continuous, such as the performance).
Both facets -- partialness and continuousness -- replace an ``is correct'' binary predicate by a ``more correct than'' relation.
In other terms, there is no direct induction between ``local binary correctness" and ``global binary correctness''.
  
This broadening of software correctness has a direct impact on software repair. One may accept a fix that partially solves a bug or that partially breaks the existing behavior. 
Those new dimensions of software correctness have been called ``acceptability envelope'' and ``approximate correctness'' by Rinard et al. \cite{rinard2005exploring}, ``controlled uncertainty'' by Locasto et al. \cite{Locasto2006} and ``sufficient correctness'' by Shaw \cite{Shaw2002}. 
However, both classical binary correctness and those unconventional kinds of correctness share a common characteristic with respect to repair:
they all implicitly define fix acceptability.

\begin{framed}
A fix is acceptable if the system stays in the correctness envelope.
\end{framed}

This perspective also gives a new light on the evaluation criteria discussed in Section~\ref{sec:eval}: the definition of the correctness envelop defines the evaluation criterion.
This point has been extensively exploited by Rinard and colleagues, for instance to automatically fix security bugs \cite{Long2012} or quality-of-service bugs \cite{misailovic2010quality}.
Note that this perspective is independent from whether repair is state-based of behavior based, and whether it happens online or offline.

\end{newtext}

\section{Related Work}
\label{sec:rw}

We structure the discussion of the related work on the following points;
the notion of defect class,
the evaluation and the risk of fallacy,
and the problem statement of automatic repair.

\subsection{On Defect Classes}
The notion of ``defect class'' or ``fault class'' is really important in the field of fault tolerance and software testing.
In fault tolerance, according to foundations of the field \cite{avizienis2004basic}, \emph{``the dependability \& security specification of a system must include the requirements [...] for specified classes of faults''}.
Indeed, one is tolerant with respect to a certain class of fault. 
Avizienis et al.'s elementary dimensions of fault classes \cite{avizienis2004basic} provide a coarse-grain framework for characterizing the bugs addressed by an automatic repair approach. 

In software testing, and in particular in the field of mutation testing, a ``fault class'' or ``fault model'' describes kinds of 
programmer mistakes (in a particular language, domain, etc) \cite{jia2011analysis}. Each mutation operator is intended to simulate one of those mistakes.
Tolerating bugs, simulating faults, repairing bugs: in all cases, there is a real need to describe the classes of bugs that are handled by a novel technique. 

 In the research on self-healing software, which is close to automatic software repair, the need for a fault model has been clearly stated by Koopman \cite{koopman2003elements}: \emph{``self-healing systems must have a fault model in terms of what injuries (faults) they are expected to be able to self-heal. Without a fault model, there is no way to assess whether a system actually can heal itself in situations of interest. ''}..

Let us now analyze other recent papers on automatic repair under the perspective of the addressed defect class.
Semfix \cite{nguyen13} is an automatic repair approach by Nguyen and colleagues based on symbolic execution. As in PAR's paper, there is no clearly addressed defect class. As said above and summarized in Table \ref{tab:defect-classes}, a defect class can be defined in terms of causes, symptoms or kinds of fix. With respect to the last point (kinds of fix), Semfix targets two clear defect classes: it fixes faulty integer initialization and faulty conditionals that use arithmetic, relational and boolean operators.
Carzaniga et al. \cite{Carzaniga2013} proposed a repair approach at runtime. In this paper, the addressed defect class is very clear, it is unhandled exceptions. 
Hosek and Cadar \cite{hosek2013safe} also address a defect class at runtime: segmentation faults (as defined by the reception of Unix' SIGSEGV signal).
The paper of Logozzo and Ball's paper \cite{logozzo2012modular} clearly conveys the notion of defect classes, up to its structure  (e.g. the section entitled \emph{``Repair of Initialization and Off-By-One Errors''}). However, they repair statically generated warnings, which are ``virtual'' bugs and not real ones.
Although those papers address defect classes, it is not always explicit.
We note that having an underlying defect class does not remove the risk of evaluation fallacy if the evaluation is conducted on a biased dataset (with respect to the defect class) or against a inappropriate competitor (idem). 

Compared to this related work, our paper explicitly states the importance of defect classes in automatic software repair.

\subsection{On Fallacies in Software Engineering}
Having sound evaluation methods is essential for science. In many fields, different fallacies have been described (e.g. \cite{goodman1999toward} in medicine). In software engineering, many authors discussed potential fallacies, such as Glass in his book \cite{glass2002facts}.
Recently, Bird and colleagues \cite{Bird2009} have extensively discussed the biases of datasets used in bug predication research. Posnett, Filkov and Devanbu \cite{posnettinference} have published a paper on the presence of ecological fallacies in empirical software engineering research (focusing on  sample size, zonation, and class imbalance). Both papers discuss the intimate relation between the dataset construction and the conclusiveness of the evaluation. 

Our paper makes the same point in a different context, automatic software repair.

\subsection{On The Problem Statement of Automatic Software Repair}

Along the two dimensions of state repair and runtime repair, let us now survey important related work.
As early as 1980, Taylor and colleagues \cite{taylor1980redundancy} introduced ``robust data structures'' which are able to repair their own state at runtime. Demsky and Rinard  \cite{demsky2003automatic} proposed a similar approach for data structure repair \cite{demsky2003automatic}, 
Perkins et al. \cite{Perkins2009} invented a complex repair strategies for register values and memory locations of x86 binary programs \cite{Perkins2009}, Friedrich et al.  \cite{Friedrich2010} focused on repairing service-oriented software. 
Lewis and Whitehead's paper  \cite{Lewis2011} also performs state repair, by runtime modification of the state of event-driven programs.

On behavioral repair, beyond the now classical work by Weimer and colleagues  \cite{Weimer2009,Goues2012}, 
there is also earlier (e.g. \cite{Jobstmann2006,Arcuri2008}) and concurrent work on this topic (e.g. \cite{Dallmeier2009,Wei2010}).
Those contributions focus on synthesizing source code to fix bugs, the code being meant to be committed into a version control system. 
However, behavioral repair is also relevant on binary code \cite{Schulte2010}.
Moreover, as stated above, behavioral repair can also happen at runtime, 
the application communities of Locasto and colleagues \cite{Locasto2006}, for instance,  share behavioral patches at runtime for fixing faults.

\section{Conclusion}
\label{sec:conclusion}

Automatic software repair is a field of research with some momentum.
It poses hard and interesting problems and may have a great impact on practitioners.
Taking as stepping stone Kim et al.'s paper published at ICSE 2013 on this topic, we have discussed the foundations of automatic repair.

First, a meaningful evaluation in automatic software repair requires one to identify and characterize a defect class.
Otherwise, there is a great risk of stating a fallacy.
Second, the apparently harmless question ``is the synthesized patch correct?'' has actually deep roots on how to define the problem statements of automatic repair and how to set up evaluation criteria.

\begin{newtext}
We are only at the beginning of automatic software repair. We are yet only able to automatically repair some bugs, in some contexts where it is easy to have a well-formed problem statement (e.g. test-suite based program repair). 
But let us open any issue tracker: how many issues can be fixed in an automated manner?
Let us have a look at any bug fix of less than 10 lines in a source code repository, for instance on Github.
How many of those small changes can be synthesized in an automated manner?
I would say very few. 
There are great inventions to be done on defining bug oracles, reproducing field failures, guiding the search for a correct repair solution, assessing the impact of synthesized changes for repair, etc..
And all this has to be done  at the scale and complexity of today's software.
\end{newtext}

\section{Acknowledgments}
First, I would like to thank Dongsun Kim, Jaechang Nam, Jaewoo Song, and Sunghun Kim, the authors of ``Automatic Patch Generation Learned from Human-Written Pat\-ches'' for giving me the opportunity and inspiration to think on such interesting problems. 
In particular, I acknowledge Dongsun Kim for his input on PAR and his insightful opinion of those critical ideas: this is a scientific debate in a very noble form.
Then, I do thank Matias Martinez for his thorough analysis of PAR and its bug evaluation dataset.
Finally, I wish to express my gratitude to Raphael Marvie, Favio DeMarco, Tegawendé F. Bissyandé, Jifeng Xuan, Lionel Seinturier, Benoit Baudry, Jean-Marc Jézéquel, Yves Le Traon, Friedrich Steimann, and Bertrand Meyer for their valuable feedback on those ideas or their deeply motivational encouragement.

This research is done with support from the Erasmus Mundus Program, the University of Lille PhD Program and EU Project Diversify FP7-ICT-2011-9.

\balance
\bibliographystyle{abbrv}
\bibliography{biblio-software-repair}

\begin{thebibliography}{10}

\bibitem{abbott1990resourceful}
R.~J. Abbott.
\newblock Resourceful systems for fault tolerance, reliability, and safety.
\newblock {\em ACM Computing Surveys (CSUR)}, 22(1):35--68, 1990.

\bibitem{ammann88}
P.~Ammann and J.~Knight.
\newblock Data diversity: an approach to software fault tolerance.
\newblock {\em IEEE Transactions on Computers}, 37(4):418 --425, 1988.

\bibitem{Arcuri2008}
A.~Arcuri and X.~Yao.
\newblock A novel co-evolutionary approach to automatic software bug fixing.
\newblock In {\em Proceedings of the IEEE Congress on Evolutionary Computation
  (CEC)}, 2008.

\bibitem{avizienis2004basic}
A.~Avizienis, J.-C. Laprie, B.~Randell, and C.~Landwehr.
\newblock Basic concepts and taxonomy of dependable and secure computing.
\newblock {\em IEEE Transactions on Dependable and Secure Computing},
  1(1):11--33, 2004.

\bibitem{Bird2009}
C.~Bird, A.~Bachmann, E.~Aune, J.~Duffy, A.~Bernstein, V.~Filkov, and
  P.~Devanbu.
\newblock Fair and balanced?: bias in bug-fix datasets.
\newblock In {\em Proceedings of the 7th joint meeting of the European Software
  Engineering Conference and the ACM SIGSOFT Symposium on the Foundations of
  Software Engineering}, 2009.

\bibitem{Carzaniga2013}
A.~Carzaniga, A.~Gorla, A.~Mattavelli, N.~Perino, and M.~Pezzè.
\newblock Automatic recovery from runtime failures.
\newblock In {\em Proceedings of the 2013 International Conference on Software
  Engineering}, 2013.

\bibitem{Dallmeier2009}
V.~Dallmeier, A.~Zeller, and B.~Meyer.
\newblock Generating fixes from object behavior anomalies.
\newblock In {\em Proceedings of the International Conference on Automated
  Software Engineering}, 2009.

\bibitem{demarco2014}
F.~DeMarco, J.~Xuan, D.~{Le Berre}, and M.~Monperrus.
\newblock Automatic repair of buggy if conditions and missing preconditions
  with {SMT}.
\newblock In {\em Proceedings of the 6th Workshop on Constraints in Software
  Testing, Verification, and Analysis Co-located with ICSE 2014}, 2014.

\bibitem{demsky2003automatic}
B.~Demsky and M.~Rinard.
\newblock Automatic detection and repair of errors in data structures.
\newblock In {\em Proceedings of the ACM SIGPLAN conference on Object-Oriented
  Programing, Systems, Languages, and Applications (OOPSLA)}, 2003.

\bibitem{Denning1989}
P.~J. Denning.
\newblock A debate on teaching computing science.
\newblock {\em Commun. ACM}, 32(12):1397--1414, Dec. 1989.

\bibitem{Friedrich2010}
G.~Friedrich, M.~Fugini, E.~Mussi, B.~Pernici, and G.~Tagni.
\newblock Exception handling for repair in service-based processes.
\newblock {\em IEEE Transactions on Software Engineering}, 36(2):198--215,
  2010.

\bibitem{glass2002facts}
R.~L. Glass.
\newblock {\em Facts and fallacies of software engineering}.
\newblock Addison-Wesley Professional, 2002.

\bibitem{goodman1999toward}
S.~N. Goodman.
\newblock Toward evidence-based medical statistics. 1: The p value fallacy.
\newblock {\em Annals of internal medicine}, 130(12):995--1004, 1999.

\bibitem{Goues2013}
C.~Goues, S.~Forrest, and W.~Weimer.
\newblock Current challenges in automatic software repair.
\newblock {\em Software Quality Control}, 21(3), 2013.

\bibitem{Goues2012}
C.~L. Goues, T.~Nguyen, S.~Forrest, and W.~Weimer.
\newblock Genprog: A generic method for automatic software repair.
\newblock {\em IEEE Transactions on Software Engineering}, 38:54--72, 2012.

\bibitem{hosek2013safe}
P.~Hosek and C.~Cadar.
\newblock Safe software updates via multi-version execution.
\newblock In {\em Proceedings of the 2013 International Conference on Software
  Engineering}, pages 612--621. IEEE Press, 2013.

\bibitem{jia2011analysis}
Y.~Jia and M.~Harman.
\newblock An analysis and survey of the development of mutation testing.
\newblock {\em IEEE Transactions on Software Engineering}, 37(5):649--678,
  2011.

\bibitem{Jobstmann2006}
B.~Jobstmann, S.~Staber, A.~Griesmayer, and R.~Bloem.
\newblock Finding and fixing faults.
\newblock {\em Lecture Notes in Computer Science}, 3725, 2005.

\bibitem{Kim2013}
D.~Kim, J.~Nam, J.~Song, and S.~Kim.
\newblock Automatic patch generation learned from human-written patches.
\newblock In {\em Proceedings of the International Conference on Software
  Engineering}, 2013.

\bibitem{koopman2003elements}
P.~Koopman.
\newblock Elements of the self-healing system problem space.
\newblock Technical report, Carnegie Mellon University, 2003.

\bibitem{Lewis2011}
C.~Lewis and J.~Whitehead.
\newblock Repairing games at runtime or, how we learned to stop worrying and
  love emergence.
\newblock {\em IEEE Software}, 28(5), 2011.

\bibitem{Locasto2006}
M.~E. Locasto, S.~Sidiroglou, and A.~D. Keromytis.
\newblock Software self-healing using collaborative application communities.
\newblock In {\em Proceedings of the 2006 Network and Distributed System
  Security Symposium (NDSS)}, 2006.

\bibitem{logozzo2012modular}
F.~Logozzo and T.~Ball.
\newblock {Modular and verified automatic program repair}.
\newblock In {\em {Proceedings of the ACM International Conference on
  Object-oriented Programming Systems, Languages and Applications}}, pages
  133--146. ACM, 2012.

\bibitem{Long2012}
F.~Long, V.~Ganesh, M.~Carbin, S.~Sidiroglou, and M.~Rinard.
\newblock Automatic input rectification.
\newblock In {\em Proceedings of the 2012 International Conference on Software
  Engineering}, 2012.

\bibitem{Martinez2013}
M.~Martinez and M.~Monperrus.
\newblock Mining software repair models for reasoning on the search space of
  automated program fixing.
\newblock {\em Empirical Software Engineering}, -, 2013.

\bibitem{Martinez2014}
M.~Martinez, W.~Weimer, and M.~Monperrus.
\newblock Do the fix ingredients already exist? an empirical inquiry into the
  redundancy assumptions of program repair approaches.
\newblock In {\em Proceedings of the International Conference on Software
  Engineering, New Ideas and Emerging Results Track (NIER)}, 2014.

\bibitem{misailovic2010quality}
S.~Misailovic, S.~Sidiroglou, H.~Hoffmann, and M.~Rinard.
\newblock Quality of service profiling.
\newblock In {\em Proceedings of the 32nd ACM/IEEE International Conference on
  Software Engineering}. ACM, 2010.

\bibitem{nguyen13}
H.~D.~T. Nguyen, D.~Qi, A.~Roychoudhury, , and S.~Chandra.
\newblock {SemFix: Program Repair via Semantic Analysis}.
\newblock In {\em Proceedings of the International Conference on Software
  Engineering}, 2013.

\bibitem{Perkins2009}
J.~H. Perkins, G.~Sullivan, W.-F. Wong, Y.~Zibin, M.~D. Ernst, M.~Rinard,
  S.~Kim, S.~Larsen, S.~Amarasinghe, J.~Bachrach, M.~Carbin, C.~Pacheco,
  F.~Sherwood, and S.~Sidiroglou.
\newblock {Automatically patching errors in deployed software}.
\newblock {\em Proceedings of the 22nd Symposium on Operating Systems
  Principles (SOSP)}, 2009.

\bibitem{posnettinference}
D.~Posnett, V.~Filkov, and P.~T. Devanbu.
\newblock Ecological inference in empirical software engineering.
\newblock In {\em Proceedings of the IEEE/ACM International Conference on
  Automated Software Engineering}, pages 362--371, 2011.

\bibitem{rinard2005exploring}
M.~Rinard, C.~Cadar, and H.~H. Nguyen.
\newblock {Exploring the acceptability envelope}.
\newblock In {\em Companion to the 20th annual ACM SIGPLAN conference on
  Object-oriented programming, systems, languages, and applications}, pages
  21--30. ACM, 2005.

\bibitem{Schulte2010}
E.~Schulte, S.~Forrest, and W.~Weimer.
\newblock Automated program repair through the evolution of assembly code.
\newblock In {\em {Proceedings of the IEEE/ACM International Conference on
  Automated Software Engineering}}, 2010.

\bibitem{Shaw2002}
M.~Shaw.
\newblock Self-healing: softening precision to avoid brittleness.
\newblock In {\em Proceedings of the first workshop on Self-healing systems},
  WOSS '02, 2002.

\bibitem{staats2011programs}
M.~Staats, M.~W. Whalen, and M.~P.~E. Heimdahl.
\newblock Programs, tests, and oracles: the foundations of testing revisited.
\newblock In {\em Proceedings of the International Conference on Software
  Engineering}, pages 391--400. IEEE, 2011.

\bibitem{taylor1980redundancy}
D.~J. Taylor, D.~E. Morgan, and J.~P. Black.
\newblock Redundancy in data structures: Improving software fault tolerance.
\newblock {\em IEEE Transactions on Software Engineering}, (6):585--594, 1980.

\bibitem{Wei2010}
Y.~Wei, Y.~Pei, C.~A. Furia, L.~S. Silva, S.~Buchholz, B.~Meyer, and A.~Zeller.
\newblock Automated fixing of programs with contracts.
\newblock In {\em Proceedings of the International Symposium on Software
  Testing and Analysis}. AC, 2010.

\bibitem{Weimer2009}
W.~Weimer, T.~Nguyen, C.~L. Goues, and S.~Forrest.
\newblock Automatically finding patches using genetic programming.
\newblock In {\em Proceedings of the International Conference on Software
  Engineering}, 2009.

\bibitem{zeller2001automated}
A.~Zeller.
\newblock Automated debugging: Are we close?
\newblock {\em Computer}, 34(11):26--31, 2001.

\end{thebibliography}

\end{document}